\documentclass[12pt]{article}
\pdfoutput=1

\usepackage{amsmath, amsfonts, amssymb}
\usepackage{graphicx}
\usepackage{psfrag}
\usepackage[usenames,dvipsnames,svgnames,table]{xcolor}
\usepackage{enumerate}
\usepackage{jheppubcustom}
\usepackage{soul}
\usepackage{subfig}

\newcommand{\be}{\begin{equation}}
\newcommand{\ee}{\end{equation}}

\def\be{\begin{equation}}
\def\ee{\end{equation}}
\def\beq{\begin{equation}}
\def\eeq{\end{equation}}

\newcommand{\del}{\delta}
\newcommand{\Del}{\Delta}

\newcommand{\de}{\partial}

\newcommand{\nab}{\nabla}

\newcommand{\avg}[1]{\langle #1 \rangle}

\renewcommand{\[}{\left[}
\renewcommand{\]}{\right]}
\renewcommand{\(}{\left(}
\renewcommand{\)}{\right)}

\renewcommand{\k}{\vec{k}}

\newcommand{\R}{\mathcal{R}}

\def\simleq{\; \raise0.3ex\hbox{$<$\kern-0.75em
      \raise-1.1ex\hbox{$\sim$}}\; }
   \def\simgeq{\; \raise0.3ex\hbox{$>$\kern-0.75em
      \raise-1.1ex\hbox{$\sim$}}\; }
      
      \newcommand{\figref}[1]{Fig. \ref{#1}}

\newcommand{\secref}[1]{Sec. \ref{#1}}

\title{Large-Scale Anomalies from Primordial Dissipation}
\author[\bigstar]{Guido D'Amico,}
\author[\bigstar]{Roberto Gobbetti,}
\author[\bigstar,\heartsuit]{Matthew Kleban,}
\author[\bigstar]{and Marjorie Schillo}

\emailAdd{gda2@nyu.edu, rg1509@nyu.edu, mk161@nyu.edu, mls604@nyu.edu}

\affiliation[\bigstar]{\it Center for Cosmology and Particle Physics, 
New York University, New York, NY  }
\affiliation[\heartsuit]{\it Institute for Advanced Study, 
Princeton, NJ  }

\begin{document}

\abstract{We analyze an inflationary model in which part of the power in density perturbations arises due to particle production.  The amount of particle production is modulated by an auxiliary field.  Given an initial gradient for the auxiliary field, this model produces a hemispherical power asymmetry and a suppression of power at low multipoles similar to those observed by WMAP and Planck in the CMB temperature.   It also predicts an additive contribution to $\delta T$ with support only at very small $l$ that is aligned with the direction of the power asymmetry and has a definite sign, as well as small oscillations in the power spectrum at all $l$.}

\maketitle

\section{Introduction}

Writ large, the data recently released by the Planck collaboration~\cite{Ade:2013zuv} are a spectacular confirmation of the standard cosmological model.  There are, however, several anomalies in the large angular scale data~\cite{Ade:2013nlj} that appear to confirm features previously observed by the WMAP satellite \cite{Eriksen:2003db, Eriksen:2007pc, Spergel:2003cb} .   Planck's confirmation of these features makes it significantly less likely that they are  systematic errors.   In this note we will present an inflationary model that attempts to account for two anomalies simultaneously:  the hemispherical power asymmetry, and the lack of power at low multipoles $l \lesssim 50$.  

It is important to note that these anomalies have most of their weight at low $l$,\footnote{Planck~\cite{Ade:2013nlj} reported a power asymmetry at high $l$ as well, but its existence is disputed \cite{FK, AL}.} and are therefore subject to a relatively large degree of uncertainty from cosmic variance.  Furthermore, they were discovered in data 
by \emph{a posteriori} analysis rather than as a prediction of a pre-existing model.  As such their true statistical significance is very difficult to estimate, and  it  remains possible -- perhaps even probable -- that these features are simply random fluctuations \cite{Bennett:2010jb}.

On the other hand, large angular scales are special:  they correspond to the earliest accessible period of inflation.  If inflation was relatively short (as there are some reasons to believe, see \emph{e.g.} \cite{Freivogel:2005vv}), remnants of the initial state may be imprinted on large scales. Observing the initial state could have a tremendous payoff as a probe of fundamental physics far beyond the reach of any earth-based experiment.  

However, even given the freedom to use any initial condition one likes, it turns out to be surprisingly difficult to account for these two anomalies with inflation.  
The reason is that inflation is exponentially efficient at erasing anisotropies in the initial state.  For effects of the initial state to be visible inflation must have been short, not lasting much longer than the minimal number ($N \sim 60$) of efolds necessary to solve the horizon and flatness problems.  

Another difficulty specific to the power asymmetry is that initial anisotropies generically affect both the fluctuations and the background  itself~\cite{Erickcek:2008sm}.  Naively, one could  consider an initial-state gradient in the inflaton itself, or in a field that couples to the inflaton and modulates the amplitude of its fluctuations.  The hemispherical power asymmetry can be described as a modulation of the power in fluctuations $\sim 10 \%$.
On the other hand $\delta T/T \sim 10^{-5}$ for all $l>1$, so any modulation of the background by the initial anisotropy must be much smaller than $10\%$.

This difficulty can be circumvented if density perturbations are not seeded (entirely) by inflaton fluctuations, but instead by fluctuations in a ``curvaton'' field that does not directly couple to the inflaton.  This was the idea pursued in~\cite{Erickcek:2009at}.  While that model is in  tension with Planck constraints, some parts of its parameter space may still be viable~\cite{Dai:2013kfa, Hirata:2013sb}.  

Here we will take a different approach.  In models of ``dissipative'' inflation, some of the fluctuations in the inflaton arise due to variations in the amount of dissipation~\cite{Green:2009ds, LopezNacir:2011kk}.  For example, in the recently proposed model of unwinding inflation~\cite{DAmico:2012sz, DAmico:2012ji}, bursts of string production occur at periodic intervals.  These strings are not produced perfectly uniformly, and the variations in their density contribute to the power spectrum of inflaton fluctuations.  
An initial gradient in a field that modulates the amount of particle production (such fields are present in all versions of unwinding inflation) will then create a hemispherical power asymmetry.  

If this modulating field is massless, its gradient is frozen during inflation and the asymmetry will be independent of $l$, which is in conflict with the result of~\cite{Hirata:2009ar} and probably with the CMB at $l \simgeq 50$ as well.  If on the other hand the modulating field is massive, its gradient will decrease at a rate determined by the mass.  For a mass of order Hubble the effect will fall off rapidly at $l \simgeq 50$.  In this case the full-sky average power in fluctuations is suppressed for $l \lesssim 50$ relative to higher $l$.

\subsection{Summary and results}

\begin{figure}
\centering
\includegraphics[scale=.5]{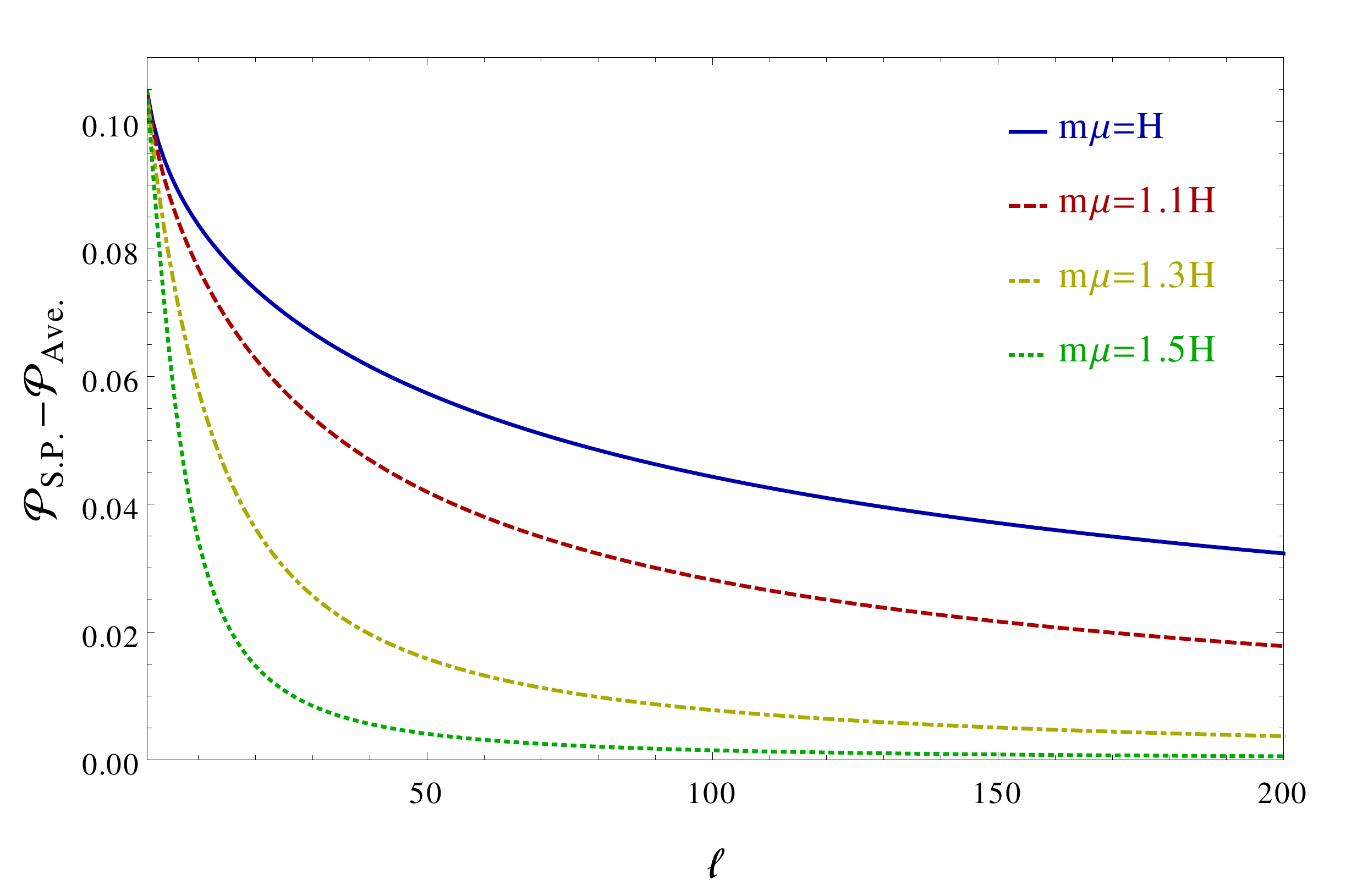}
\caption{Suppression of average power at low $l$ plotted for different values of $m_\mu$.}
\label{fig:suppression}
\end{figure}

\begin{figure}
\centering
\includegraphics[scale=.5]{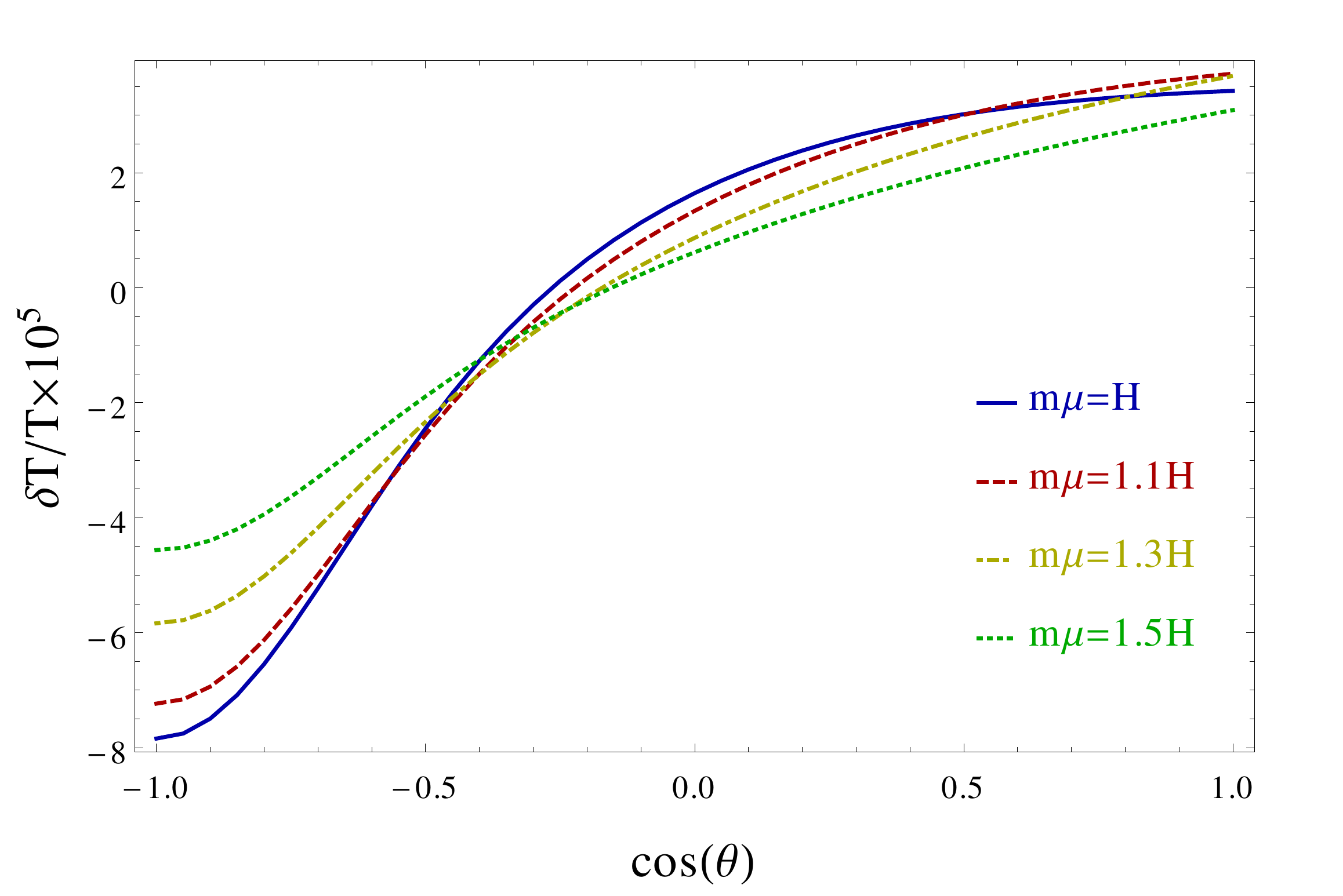}
\caption{Additive temperature anomaly $\delta T(\theta)$ on the CMB sky due to the variation in the amount of particle production, for different values of the mass of $\mu$.  The direction is aligned with the power asymmetry, and the temperature \emph{increases} with \emph{decreasing} power in fluctuations. }
\label{fig:temperature}
\end{figure}

We will present a model with the following characteristics:

\begin{itemize}
\item Inflationary perturbations arise from two sources: de Sitter fluctuations of the inflaton $\phi$ ($\sim90 \%$ of the power) and fluctuations due to particle production  ($\sim10 \%$ of the power) 
\item A field $\mu$ that modulates the amount of particle production, and initially ($\sim 60$ efolds from the end of inflation) has a long wavelength gradient in direction $\hat d$
\item The gradient reduces almost to zero after $\sim 4$ efolds of inflation because $\mu$ has a mass $m_\mu \gtrsim H$.
\end{itemize}
The predictions of this model are:

\begin{itemize}
\item A hemispherical power asymmetry $\sim10 \%$ in direction $\hat d$ for $l \simleq 50$ (\figref{fig:difference})
\item A suppression in average low-$l$ power across the entire sky of $\sim5 \%$ for $l \simleq 50$ (\figref{fig:suppression})
\item An additive, deterministic temperature anomaly $\delta T(\hat n \cdot \hat d)$ aligned with the power asymmetry and affecting multipoles with $l \simleq 7$ (\figref{fig:temperature}, \figref{fig:Tsphere})
\item Small oscillations in the power spectrum at all $l$.
\end{itemize}

\subsection{Previous work}

A recent discussion of the power asymmetry can be found in \cite{Dai:2013kfa}.  There have been a number of previous attempts to explain the asymmetry with inflationary dynamics, either by introducing a curvaton or with scale-dependent non-Gaussianity  \cite{Erickcek:2008sm, Erickcek:2009at, Lyth:2013vha}.  These models are in some tension with data, although the model of  \cite{Erickcek:2009at} may still be allowed  \cite{Dai:2013kfa}.  After this work was completed \cite{Liddle:2013rta} appeared, which extends the model of \cite{Erickcek:2008sm} to open universes.

\section{Asymmetry from particle production}

The model we will study is a simplified version of the low-energy effective description of unwinding inflation \cite{DAmico:2012sz,DAmico:2012ji}, and is very similar to trapped inflation \cite{Green:2009ds}.  In unwdinding inflation the kinetic terms for fields are of DBI type \cite{Alishahiha:2004eh}, there are multiple string modes (including one that is briefly tachyonic) produced periodically throughout inflation when a brane and anti-brane pass close to each other, and there are several light scalars that modulate the mass of the produced particles/strings.\footnote{Another interesting characteristic of unwinding inflation is that it is initiated by bubble nucleation, and therefore the spatial curvature of the universe is open.}  The latter arise because there are ``impact parameters"---the relative separation of the brane and the anti-brane in the compact dimensions transverse to them.
For now we will ignore the DBI kinetic terms (see \secref{DBI}) and focus on a toy model in which there is one inflaton $\phi$, only one transverse scalar $\mu$ and zero bare mass for the produced scalar particles $\chi_i$.

Consider the action
\begin{multline}
\label{eq:action}
S = - \int \sqrt{-g} \bigg\{ \frac{1}{2} (\de \phi)^2 + {1 \over 2} m^2 \phi^2
+ \frac{1}{2} (\de \mu)^2 + \frac{1}{2}m_\mu^2 \mu^2 \\
+ \sum_i \[\frac{1}{2} (\de \chi_i)^2 + \frac{g^2}{2} \(\mu^2 + (\phi - \phi_i)^2 \) \chi_i^2\] 
\bigg\} \, ,
\end{multline}
where $\phi_i = i \Delta \phi$ (with $i$  integer) are evenly spaced points in $\phi$-space.

This action describes a slow-rolling inflaton $\phi$ coupled to a sector of fields $\chi_i$, each of which passes through a point of minimum mass when  $\phi = \phi_i$.  For slow roll, $\phi(t) \approx \dot \phi t$ and these intervals are equally spaced in physical time.  Since the mass of $\chi_i$ is time-dependent, $\chi_i$ particles may be produced spontaneously at the times when $\phi \approx \phi_i$.  Because of the $\mu^2 \chi_i^2$ coupling in~\eqref{eq:action}, the VEV of $\mu$ contributes to the mass of $\chi_i$
 and controls the amount of particle production.  If $\mu$ is large, $\chi_i$ is heavy even when $\phi=\phi_i$ and very few $\chi_i$ quanta are produced.  On the other hand if $\mu = 0$ the $\chi$ particles become instantaneously massless, leading to significant $\chi_i$ particle production.

Curvature perturbations are related to fluctuations in the inflaton $\delta \phi$ as usual, by $\R =  - H \delta \phi/\dot \phi$.  The novel feature is that the fluctuations $\delta \phi$ arise due to a combination of standard de Sitter effects and variations in the density of produced particles (\cite{Green:2009ds, LopezNacir:2011kk, DAmico:2012ji}).  If $\mu \sim 0$, production of $\chi_i$ particles is unsuppressed and contributes to perturbations $\delta \phi$.  On the other hand, if $\mu$ is large particle production is suppressed, so that  only de Sitter fluctuations contribute to $\delta \phi$. 

Suppose that  at the start of inflation at time $t=t_i$, $\mu$ has a linear gradient in direction $\hat d$:
\be \label{mugrad}
\mu(\vec x, t_i) = \mu_0 + \alpha ( \hat d \cdot  \vec x + R),
\ee
where $R$ is the radius of the last scattering surface.  In this case particle production will be more suppressed in one part of the universe than another.  For simplicity, we will set $\mu_{0}=0$, so that $\mu=0$ at a point on the last scattering sphere.  

This  will create a power asymmetry
\be
\label{dipolea} 
\delta T (\hat n) \sim \delta T_{0} (\hat n)(1 + A  \, \hat n \cdot \hat d),
\ee
where $\delta T_{0}(\hat n)$ is  isotropic and the amplitude $A$ is at most the fraction of the power coming from particle production at $\mu=0$.
This is the phenomenological ansatz proposed in~\cite{Gordon:2005ai}.  

\begin{figure}
\begin{center}$
\begin{array}{c c }
	\includegraphics[angle=0, width=0.5\textwidth]{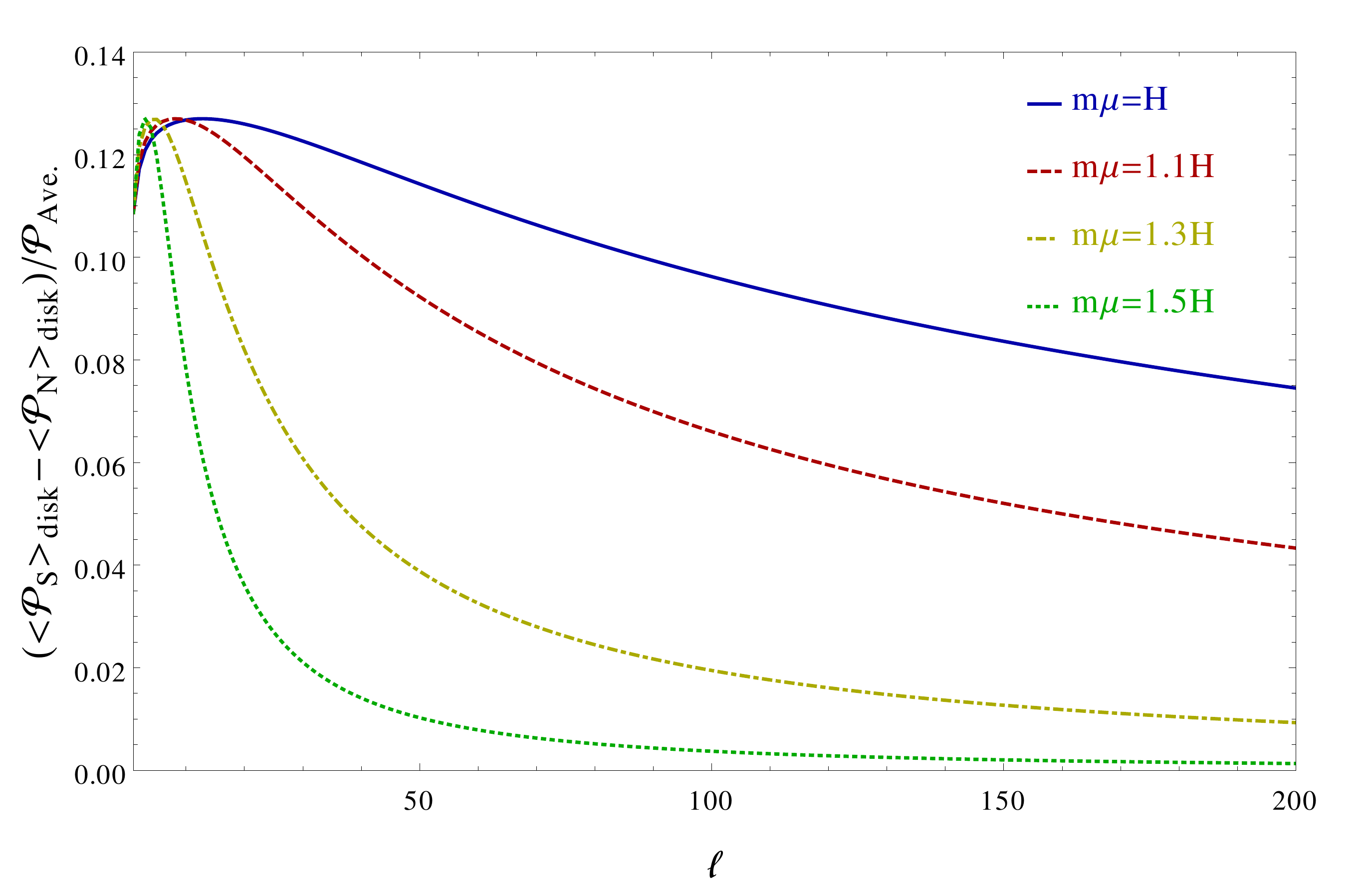} & \includegraphics[angle=0, width=0.5\textwidth]{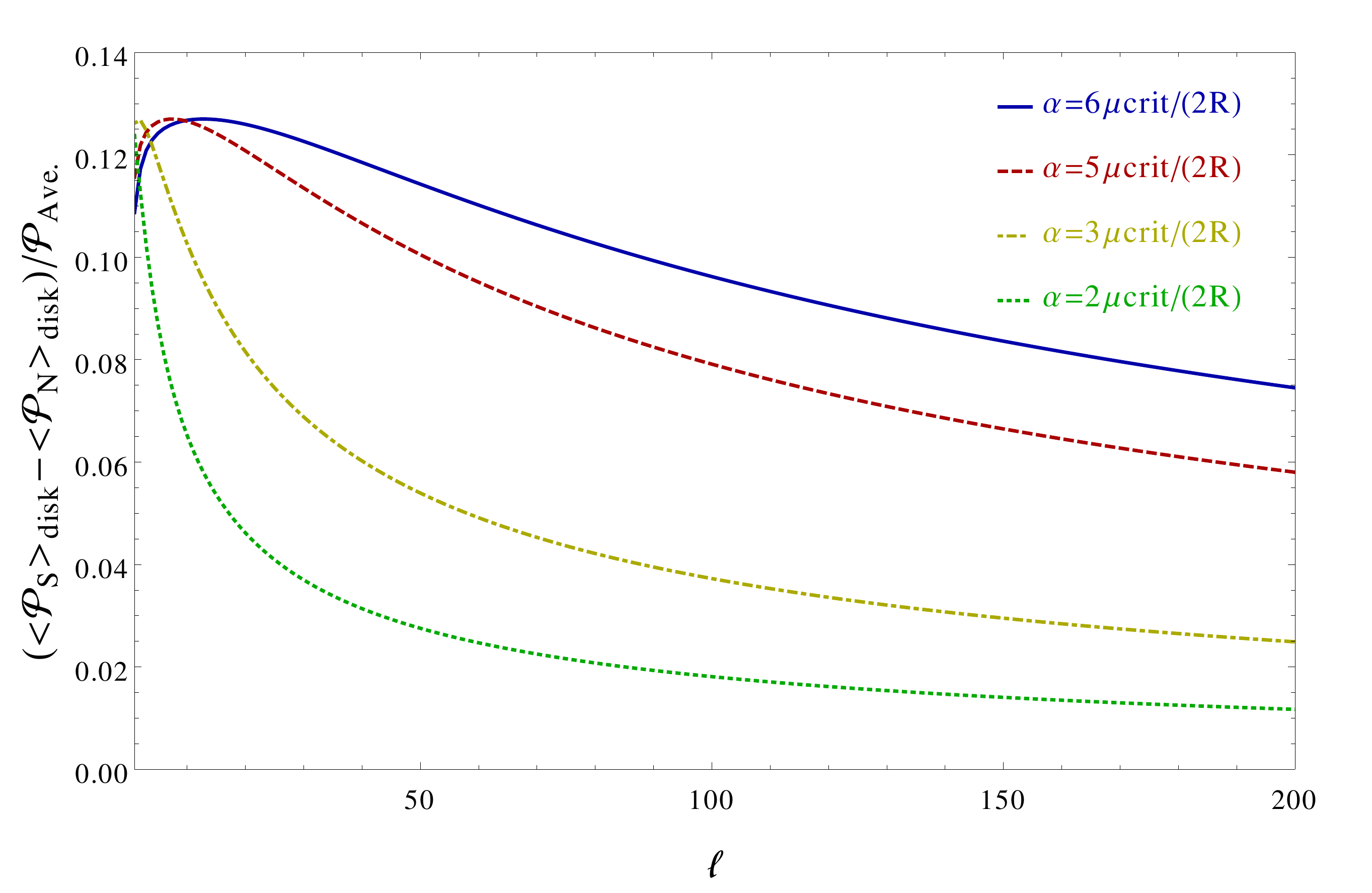} \\
\end{array}$
\end{center}
\caption{Difference in average power between a disk of $45^{\circ }$ radius centered on  the north pole (with respect to the asymmetry axis $\hat d$) and one centered on the south pole, divided by the average power.  Left panel:  $\alpha=6 \mu_{crit}/2R$ (see \eqref{mugrad}), varying  $m_\mu$.   Here $\mu_{crit}^{2} = \dot{\phi}/ \pi g$ (see \eqref{nchi} and \eqref{muc}) and $2 R$ is the diameter of the last scattering sphere.  Right panel: $m_{\mu} = H$, varying $\alpha$.}
\label{fig:difference}
\end{figure}

\begin{figure}
\centering
\includegraphics[scale=.5]{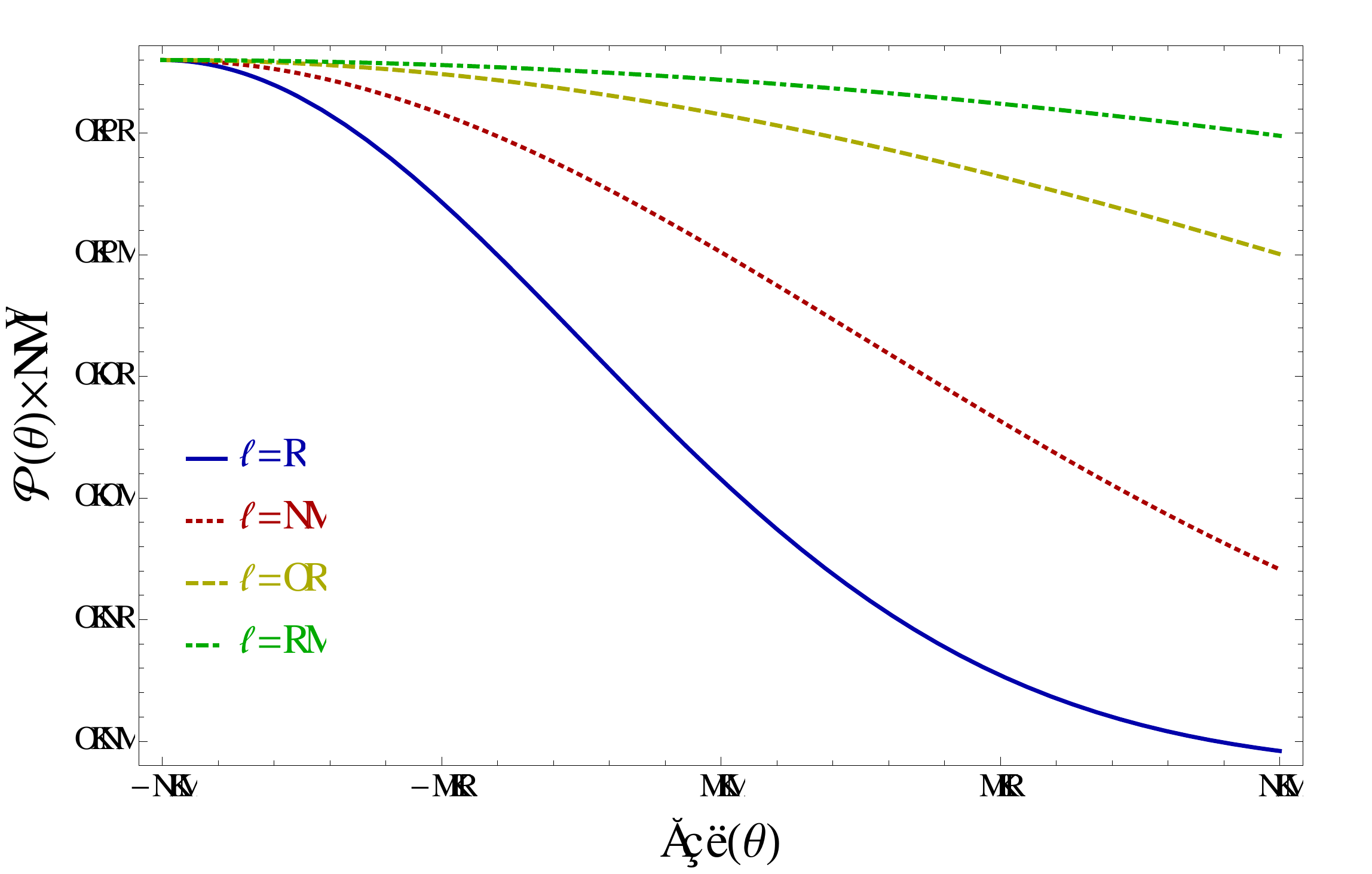}
\caption{Power spectrum in  fluctuations as a function of angle from the asymmetry direction.  The $l$ dependence arises because larger $l$ corresponds to later times during inflation, when the gradient in $\mu$ has had more time to relax.}
\label{fig:modulation}
\end{figure}

The ``$\sim$'' in \eqref{dipolea} arises because the effect of $\mu$ on the power spectrum is non-linear, and because the gradient in $\mu$ will generally evolve in time.  If $\mu$ is massless and non-interacting, its initial gradient will remain permanently imprinted and the power asymmetry would persist to high $l$.  However for $m_\mu \neq 0$, the gradient will change during inflation -- specifically, $\mu$ will eventually relax to zero everywhere and the asymmetry (and overall suppression in power) will disappear at high $l$, modifying the ansatz~\eqref{dipolea}.  
In Figure~\ref{fig:difference} we plot the power asymmetry, while Figure~\ref{fig:modulation} illustrates the $l$ dependence of $A$.

\subsection{Equations of motion}
The equations of motion derived from~\eqref{eq:action} are:
\begin{align}
&\ddot{\phi} + 3 H \dot{\phi} - \frac{\nab^2}{a^2} \phi  + m^2 \phi + g^2 \sum_i (\phi - \phi_i) \chi_i^2 = 0 \\
&\ddot{\mu} + 3 H \dot{\mu} - \frac{\nab^2}{a^2} \mu  + m_\mu^2 \mu + g^2 \sum_i \mu \chi_i^2 = 0 \\
&\ddot{\chi_i} + 3 H \dot{\chi_i} - \frac{\nab^2}{a^2} \chi_i + g^2 \sum_i \[ \mu^2 + (\phi - \phi_i)^2 \] \chi_i = 0 \, .
\end{align}
There are bursts of particle production when $\phi \approx \phi_i$ due to violation of adiabaticity.
In each burst of particle production, a comoving number density of $\chi$ particles
\be \label{nchi}
n_\chi = \frac{(g \dot{\phi})^{3/2}}{(2 \pi)^3} e^{-\pi g \mu^2 /\dot \phi}
\ee
is produced, where $g \mu$ is the minimum mass for $\chi_i$ (i.e. the mass when $\phi=\phi_i$) \cite{Kofman:2004yc}.
This formula is valid if the time interval of non-adiabaticity $\del t \sim (g \dot{\phi})^{-1/2} \ll H^{-1}$.  Note that particle production is exponentially suppressed when $\mu > \mu_{crit}$, where
\be \label{muc}
\mu_{crit} \equiv \sqrt{\dot \phi/(\pi g)}
\ee

For $\phi>\phi_i$ the mass of the $\chi$ particles increases linearly with $\phi - \phi_i$, so that their comoving energy density  is
\be
\rho_\chi = m_\chi n_\chi = g \sqrt{\mu^2 + (\phi - \phi_i)^2} n_\chi,
\ee
\cite{Green:2009ds,  DAmico:2012ji}, and their physical energy density redshifts as $a^{-3}$ from the time they are produced on.

The back-reaction of $\chi$ particle production on the background trajectory of $\phi$ and $\mu$ can be estimated as in~\cite{Kofman:1997yn,  DAmico:2012ji}.  We are interested in scenarios where $\sim 10\%$ of the power in fluctuations comes from particle production.
In this case the back-reaction on the background is small -- but potentially important nevertheless, as we shall see.

For most of the rest of this paper we will ignore the fact that particle production occurs in bursts at periodic intervals in time, and instead work with time averages.  This approximation is valid when there are many collisions per Hubble time:
$$N_{coll} \equiv \dot \phi/(H \Delta \phi) \gg 1.$$  
When this condition is not satisfied, scale-invariance will be broken by  oscillations in the amplitude of the component of the power spectrum that comes from particle production.  We will comment on this briefly in \secref{osc}.

The time-averaged background equation of motion for $\phi$ is
\be
\label{eq:infeom}
\ddot{\phi} + 3 H \dot{\phi} + m^2 \phi = \bar{f} \, ,
\ee
where we have defined the time-averaged ``force"
\be
\bar{f} = - \frac{\de \bar \rho_\chi}{\de \phi} = - \frac{\dot{\phi}}{3 H \Del \phi} \frac{g^{5/2} \dot{\phi}^{3/2}}{(2 \pi)^3} e^{-\pi g \mu^2/ \dot \phi} \, .
\ee
Here $\Del \phi$ denotes the spacing between two consecutive $\phi_i$'s, and the factor of $\dot{\phi}/(3 H \Delta \phi)$ arises from the time average.

Particle production modifies the background inflaton trajectory because of the force term $\bar f$ in \eqref{eq:infeom}.
In the presence of a large-scale spatial gradient in $\mu$, $\bar{f}$ varies across the universe, so that inflation lasts slightly longer in regions with larger particle production.  This produces a large-scale temperature variation aligned with the direction of the gradient.  We will explore this in detail in the next subsection.

\subsection{Perturbations}
As mentioned above, the production of $\chi$ particles produces a new term in the inflaton equation of motion~\eqref{eq:infeom} that affects the slow-roll velocity in a $\mu$-dependent way.   In addition, random variations in the density of produced particles act as a source for adiabatic perturbations $\delta \phi$.

The equation for the perturbations $\delta \phi$ is:
\be
\ddot{\del \phi}_{\k} + 3 H \dot{\del \phi}_{\k} + \frac{k^2}{a^2} \del \phi_{\k} - \frac{\de \bar{f}}{\de \dot{\phi}} \dot{\del \phi}_{\k} = g \bar{\del n}_{\k},
\ee
where $\bar{\del n}_{\k}$ is the spatial variation in (the time average of) the number density of produced particles.  Time averaging approximates the rate and physical number density of produced particles as constant in time (up to slow-roll corrections), guaranteeing the scale-invariance of the resulting spectrum of perturbations.  

If we 
assume particle production is a Poisson process, the variance in the number density of produced particles is 
\be
\avg{\bar{\del n}_{\k} \bar{\del n}_{\k'}} = (2 \pi)^3 \del_D(\k + \k') \frac{\bar{n}_\chi}{a^3} \, ,
\ee
where $\bar{n}_\chi = \dot{\phi} n_\chi/(3 H \, \Del \phi)$ is the time-averaged number density produced.
Assuming that particle production is uncorrelated with the de Sitter fluctuations in $\delta \phi$, the final result for the total power spectrum of the curvature perturbation is
\be
\label{eq:fullPS}
\mathcal{P}_\R = 
\mathcal{P}_{dS} + \mathcal{P}_{ pp}=\frac{H^4}{4 \pi^2 \dot{\phi}^2} +\frac{g^2 \bar{n}_\chi H}{9 \pi \dot{\phi}^2}
\ee
where $\mathcal{P}_{dS}$ is due to de Sitter fluctuations and $\mathcal{P}_{ pp}$ is  due to particle production.  
Therefore, if  $\mu$ varies  from $0$ to $\mu_{crit}$ or greater (where $n_{\chi}$ is suppressed by the exponential in \eqref{nchi} and $\mathcal{P}_{ pp} \simeq 0$), the difference in power is
\be
\label{eq:PowAsym}
\frac{\Del \mathcal{P}}{\mathcal{P}} = \frac{\mathcal{P}_{ pp}}{\mathcal{P}_{dS} + \mathcal{P}_{ pp}}
\simeq \frac{\dot{\phi}}{H \Delta \phi} \frac{g^{7/2} \dot{\phi}^{3/2}}{54 \pi^2 H^3} = N_{coll} \mathcal{P}_0^{-3/4} \frac{g^{7/2}}{108 \pi^3 \sqrt{2 \pi}}.
\ee
It is easy to arrange $\Del \mathcal{P}/\mathcal{P} \simeq 10 \%$, for instance by choosing $g = 0.07$ and $N_{coll} = 4$ (these are the values used in all plots).

Turning to the effects on the background evolution, the term $\bar f$ in \eqref{eq:infeom} causes  a \emph{deterministic}  change $\delta N$ in the number of efolds of inflation between regions with zero and non-zero initial values of $\mu$:
\be \label{dN}
\delta N \simeq \frac{\bar{f}}{3 H \dot{\phi}} \Delta N,
\ee
where $\Delta N$ is the number of efolds during which $\mu$ has a non-negligible gradient.
In turn, $\delta N$ translates into an additive temperature anomaly via the Sachs-Wolfe effect:
$$
\delta T/T \approx - \frac{1}{5} \delta N.
$$

\begin{figure}
\begin{center}$
\begin{array}{| c| c |}
\hline
	\includegraphics[angle=0, width=0.5\textwidth]{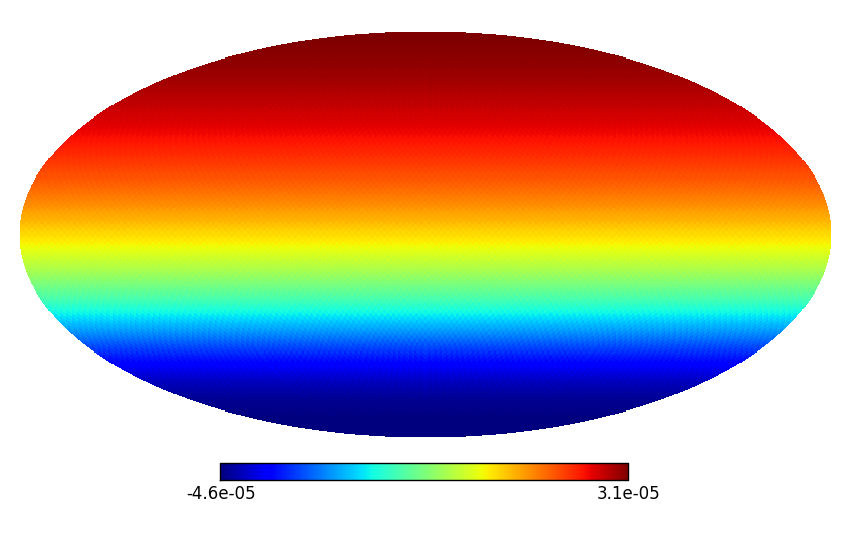} & \includegraphics[angle=0, width=0.5\textwidth]{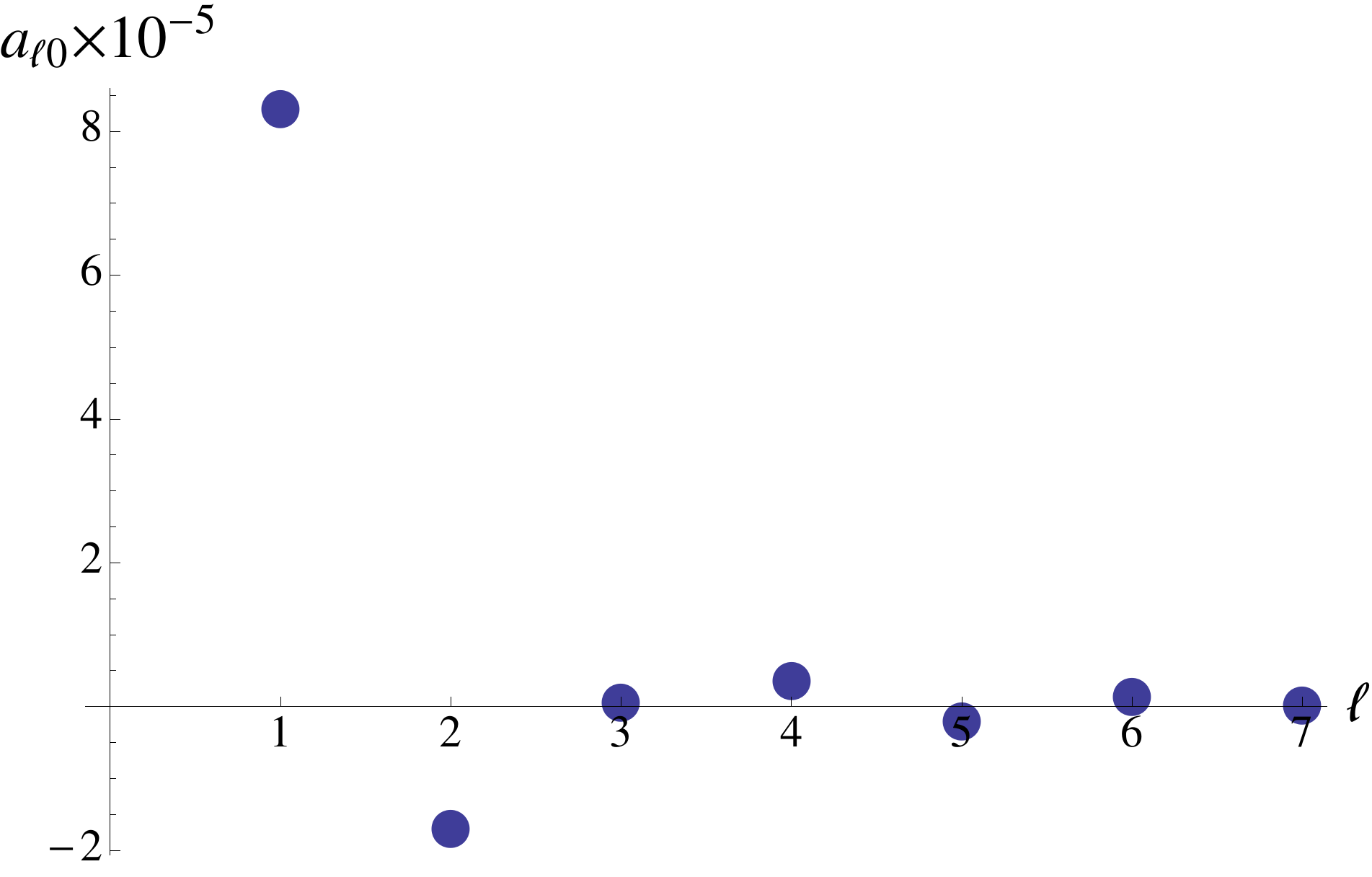} \\
		\hline
\end{array}$
\end{center}
\caption{Left panel: Temperature anomaly $\delta T(\theta)/T$ on the CMB sky due to the variation in the amount of particle production.  The direction is aligned with the power asymmetry, and the temperature \emph{increases} with \emph{decreasing} power in fluctuations.  Right panel: multipole moments $a_{l0}$ of $\delta T/T$ in a coordinate system aligned with the power asymmetry.  Both plots are with $m_\mu = 3H/2$.}
\label{fig:Tsphere}
\end{figure}

For an initially linear gradient this contribution to $\delta T$ is primarily dipole (\figref{fig:Tsphere}), but contributes to a few higher multipoles as well.  It is aligned with the direction of the power asymmetry, and scales in proportion to the strength of the asymmetry.  If the power asymmetry monotonically increases from one pole to the other, so does this temperature anomaly.  Moreover, the temperature anomaly has a definite \emph{sign}.  More particle production means more power in fluctuations and a larger friction term.  Therefore inflation lasts slightly longer in regions of the sky with higher power, and hence the temperature is lower than average  in those regions.  

To estimate the maximum size of the effect:
\be
\label{eq:FricRatio}
\frac{\bar{f}}{3 H \dot{\phi}} \simeq \frac{N_{coll}}{9 H \dot{\phi}} \frac{g^{5/2} \dot{\phi}^{3/2}}{(2 \pi)^3} = N_{coll} \mathcal{P}_{dS}^{-1/4} \frac{g^{5/2}}{9 (2\pi)^{7/2}}
= \( \frac{3 \mathcal{P}_{dS}^2}{32}\)^{1/7} \(\frac{\Del \mathcal{P}}{\mathcal{P}}\)^{5/7}  \frac{N_{coll}^{2/7}}{2\pi}
\simeq 7 N_{coll}^{2/7} \times 10^{-5} 
\ee
where we have set $\Del \mathcal{P}/\mathcal{P} = 0.1$.
The resulting temperature anomaly has amplitude
$$
\frac{\delta T}{T} \approx - \frac{7 \Delta N}{5} N_{coll}^{2/7} \times 10^{-5}
$$
where $\Delta N$ is the number of efolds during which $\mu$ varies significantly relative to $\mu_{crit} = \dot{\phi}^{1/2}/(\pi g)^{1/2}$.
If we require that the power asymmetry approaches zero at $l \sim 50$,  $\Del N \simeq \ln 50 \approx 4$.
This anomaly and its associated spherical moments are shown in \figref{fig:Tsphere}.

If this temperature anomaly is monotonically increasing from pole to pole and confined to very low $l$, a useful statistic to detect it is simply the difference in temperature between the north and south poles evaluated on a smoothed map.  This quantity receives contributions only from $m=0$, and only from odd $l$:
$$
\frac{(T_N - T_S)_{\rm smooth}}{\bar{T}} = \sum_{{\rm odd}\,l} \sqrt{\frac{2 l+1}{\pi}} a_{l0}.
$$

\subsection{Oscillations} \label{osc}

For the most of the paper, we have been working with time averages.
In this approximation, the particle production produces a scale-invariant component of the power-spectrum.
However, the particle production occurs periodically in physical time, and therefore its contribution to the power spectrum is oscillatory  in $\log k$.
The frequency is proportional to $N_{coll}$, and the amplitude suppressed by $1/N_{coll}$ to a power. In order for our time-averaging approximation to be valid, we require $N_{coll} \simgeq O(5)$.  Lower  values would produce potentially detectable oscillations in the power spectrum at high $l$, and possibly interesting features (such as dips) at lower $l$.  We will leave this possibility for future work.

\subsection{Polarization}

Any large-scale modulation of the power in primordial density fluctuations produces a correlated signal in polarization, because the quadrupole CMB anisotropy seen by free electrons at reionization will vary across the sky. According to \cite{Dvorkin:2007jp}, an approximately dipolar modulation in CMB power of the form \eqref{dipolea} can be tested at $\sim 98\%$ confidence with large-angle polarization data. Because our model produces a similar modulation at least for $l \simleq 50$, this analysis should apply to it.  However, \cite{Dvorkin:2007jp} used a dipolar modulation amplitude $A \approx 0.11$ (based on \cite{Eriksen:2007pc}, an analysis of WMAP 3-year data), whereas current analyses favor a smaller value $A\approx .07$.
As such the significance of the correlated polarization signal would be reduced.

\subsection{Relativistic kinetic terms} \label{DBI}
In unwinding inflation, the inflaton has a non-canonical, DBI-type kinetic term that limits $\dot \phi$ from exceeding a certain value.  Physically this is because $\phi$ is the position of a brane in an extra dimension, and the brane's velocity cannot exceed the speed of light.  Hence incorporating a DBI kinetic term for $\phi$ into ~\eqref{eq:action} introduces an additional parameter, the maximum value of $\dot \phi$ \footnote{There is a DBI kinetic term for $\mu$ as well, but since $\dot \mu$ remains small these would have very little effect on the dynamics.}. 

This additional degree of freedom makes it somewhat easier to match the power asymmetry anomaly while satisfying observational constraints.  In the non-relativistic case $m$ (the mass of the inflaton) and $\dot{\phi}$ are fixed by requiring that $P_0$ agree with observation, but in the relativistic case $P_0$ depends on the new parameter, as does the power asymmetry \eqref{eq:PowAsym}.  As a result one can vary $N_{coll}$ after fixing the coupling constant $g$ and still achieve the desired degree of asymmetry.  For example, in the relativistic case the deterministic perturbation to $\delta T /T$ can be reduced by adjusting the ratio of friction terms in \eqref{eq:FricRatio}.\footnote{In string theory there are many open string modes that become light when  branes pass close to each other, significantly increasing the amount of particle production relative to the model considered here.}

\section{Initial conditions}

So far, we have presented a mechanism by which an appropriately chosen initial condition results in both a power asymmetry and low power at large scales in the CMB.  But how could the  initial conditions  have come about, and how natural or likely are they?

Inflation restores isotropy with exponential efficiency, meaning that anisotropic quantities decrease exponentially with the number of efolds.  Hence, the hemispherical power asymmetry seems to require either nearly minimal inflation with appropriate initial conditions, or an event that occurred during long inflation and created anisotropy.  
If inflation was near-minimal and the initial conditions were chaotic or random, after a few efolds of inflation the wavelength of the initial state perturbations will stretch to superhorizon size.  Given an appropriate multi-field model such as the one we have presented, this can give rise to a power asymmetry, but on the other hand there is no reason for the amplitude of the power asymmetry to be significantly larger than the amplitude of the quadrupole~\cite{GZ}.

One well-studied alternative is a cosmic bubble collision (see \cite{Kleban:2011pg} for a review and further references).  The collision creates a ``cosmic wake'' \cite{Kleban:2011yc} of energy that propagates across our universe, leaving behind a perturbed region in which the fields are affected in various ways.  The wake can cross our Hubble volume either before inflation or some time after it starts.  In either case after a few efolds the dominant effect on each field is a gradient pointing towards the collision, with a slope that depends on the coupling of the field to the bubble wall. The gradient is linear plus smaller corrections, because higher powers in the Taylor expansion are suppressed by additional powers of $e^{-N}$ (see \cite{Gobbetti:2012yq} for details).  

Most of the analyses of the effects of collisions have focused on the case where the edge of the wake bisected our Hubble volume at last scattering.  For a collision with a thin-wall bubble, the edge of the wake sharply bounds the region in which there is a gradient from the one in which there is no effect, creating an edge and a set of characteristic signatures in the CMB.  However, the majority of collisions create wakes that passed across our entire Hubble volume well before last scattering.  The leading effect of these is to introduce superhorizon gradients in the various fields that are affected by the collision.  

Bubble collisions are special in that the gradients they produce in all fields are aligned, as are all the sub-leading non-linear corrections.  In other words all effects are invariant under rotations around the axis pointing towards the center of the colliding bubble.  In a coordinate system aligned with this axis, only the $a_{l0}$ spherical moments receive contributions from the collision, and (because there is no chirality) polarization is entirely $E$-mode.

Another bubble-related origin for initial anisotropy is a fluctuation away from homogeneity and isotropy -- which arises due to the spherical symmetry of the bubble nucleation instanton in Euclidean space -- in the fields in our bubble when it forms.  This could lead to large-scale initial anisotropies in $\mu$ and other fields.  However, such fluctuations are generally peaked at the length scale set by the spatial curvature, so constraints on $\Omega_k$ might make this scenario difficult to realize.\footnote{We thank P.~Creminelli  for discussions on this point.} 

\section{Conclusions}
 Ideally, attempts to explain the origin of the anomalies in the Planck data should account for more than one of them, and make correlated predictions for other quantities as well.   Here, we presented a model for inflationary dynamics that, given appropriate initial conditions, produces two anomalies: a large-scale power asymmetry and a lack of power at low multipoles.  Moreover it produces a deterministic $\delta T$ with support at very low $l$ that is aligned with the direction of the power asymmetry and has a definite sign.  We believe this is kind of aligned temperature anomaly is a generic prediction of any model that accounts for the power asymmetry by modulating the power in inflaton perturbations (although the sign is probably model-dependent).  In addition our model predicts oscillations in the power spectrum, and if it arises from unwinding inflation, predicts an open universe and a combination of $r$ and $f_{NL, eq}$ that should be confirmed or ruled out by near-future data \cite{DAmico:2012ji}.
Determining how well this model fits the Planck data requires a dedicated statistical analysis which we leave for the future.

\section*{Acknowledgements}
It is a pleasure to thank P.~Creminelli, R.~Flauger, E.~Komatsu, A.~Lewis, M.~Roberts,  L.~Senatore, and M.~Zaldarriaga for discussions.  M.K.  and G.D'A. thank the KITP Santa Barbara for hospitality during the Primordial Cosmology program.
G.D'A. is supported by a James Arthur Fellowship.
The work of MK is supported in part by the NSF  through grant PHY-1214302, and by the John Templeton Foundation.  The opinions expressed
in this publication are those of the authors and do not necessarily reflect the views of the John Templeton Foundation.

\bibliography{Bibliography}

\providecommand{\href}[2]{#2}\begingroup\raggedright\begin{thebibliography}{10}

\bibitem{Ade:2013zuv}
{\bfseries Planck} Collaboration, {\sc P.~Ade} et~al., ``{Planck 2013 results.
  XVI. Cosmological parameters},''
\href{http://arxiv.org/abs/1303.5076}{{\ttfamily arXiv:1303.5076
  [astro-ph.CO]}}.

\bibitem{Ade:2013nlj}
{\bfseries Planck} Collaboration, {\sc P.~Ade} et~al., ``{Planck 2013 results.
  XXIII. Isotropy and Statistics of the CMB},''
\href{http://arxiv.org/abs/1303.5083}{{\ttfamily arXiv:1303.5083
  [astro-ph.CO]}}.

\bibitem{Eriksen:2003db}
{\sc H.~Eriksen}, {\sc F.~Hansen}, {\sc A.~Banday}, {\sc K.~Gorski}, and {\sc
  P.~Lilje}, ``{Asymmetries in the Cosmic Microwave Background anisotropy
  field},'' \href{http://dx.doi.org/10.1086/382267}{{\em Astrophys.J.}
  {\bfseries 605} (2004) 14--20},
\href{http://arxiv.org/abs/astro-ph/0307507}{{\ttfamily arXiv:astro-ph/0307507
  [astro-ph]}}.

\bibitem{Eriksen:2007pc}
{\sc H.~K. Eriksen}, {\sc A.~Banday}, {\sc K.~Gorski}, {\sc F.~Hansen}, and
  {\sc P.~Lilje}, ``{Hemispherical power asymmetry in the three-year Wilkinson
  Microwave Anisotropy Probe sky maps},''
  \href{http://dx.doi.org/10.1086/518091}{{\em Astrophys.J.} {\bfseries 660}
  (2007) L81--L84},
\href{http://arxiv.org/abs/astro-ph/0701089}{{\ttfamily arXiv:astro-ph/0701089
  [astro-ph]}}.

\bibitem{Spergel:2003cb}
{\bfseries WMAP} Collaboration, {\sc D.~Spergel} et~al., ``{First year
  Wilkinson Microwave Anisotropy Probe (WMAP) observations: Determination of
  cosmological parameters},'' \href{http://dx.doi.org/10.1086/377226}{{\em
  Astrophys.J.Suppl.} {\bfseries 148} (2003) 175--194},
\href{http://arxiv.org/abs/astro-ph/0302209}{{\ttfamily arXiv:astro-ph/0302209
  [astro-ph]}}.

\bibitem{FK}
{\sc R.~Flauger} and {\sc E.~Komatsu} Private communication, June 2013.

\bibitem{AL}
{\sc A.~Lewis}, ``{Primordial and kinematic power modulation from Planck},''.
\newblock (Talk given at Observations and Theoretical Challenges in Primordial
  Cosmology conference, April 2013).

\bibitem{Bennett:2010jb}
{\sc C.~Bennett}, {\sc R.~Hill}, {\sc G.~Hinshaw}, {\sc D.~Larson}, {\sc
  K.~Smith}, et~al., ``{Seven-Year Wilkinson Microwave Anisotropy Probe (WMAP)
  Observations: Are There Cosmic Microwave Background Anomalies?},''
  \href{http://dx.doi.org/10.1088/0067-0049/192/2/17}{{\em Astrophys.J.Suppl.}
  {\bfseries 192} (2011) 17},
\href{http://arxiv.org/abs/1001.4758}{{\ttfamily arXiv:1001.4758
  [astro-ph.CO]}}.

\bibitem{Freivogel:2005vv}
{\sc B.~Freivogel}, {\sc M.~Kleban}, {\sc M.~Rodriguez~Martinez}, and {\sc
  L.~Susskind}, ``{Observational consequences of a landscape},''
  \href{http://dx.doi.org/10.1088/1126-6708/2006/03/039}{{\em JHEP} {\bfseries
  0603} (2006) 039},
\href{http://arxiv.org/abs/hep-th/0505232}{{\ttfamily arXiv:hep-th/0505232
  [hep-th]}}.

\bibitem{Erickcek:2008sm}
{\sc A.~L. Erickcek}, {\sc M.~Kamionkowski}, and {\sc S.~M. Carroll}, ``{A
  Hemispherical Power Asymmetry from Inflation},''
  \href{http://dx.doi.org/10.1103/PhysRevD.78.123520}{{\em Phys.Rev.}
  {\bfseries D78} (2008) 123520},
\href{http://arxiv.org/abs/0806.0377}{{\ttfamily arXiv:0806.0377 [astro-ph]}}.

\bibitem{Erickcek:2009at}
{\sc A.~L. Erickcek}, {\sc C.~M. Hirata}, and {\sc M.~Kamionkowski}, ``{A
  Scale-Dependent Power Asymmetry from Isocurvature Perturbations},''
  \href{http://dx.doi.org/10.1103/PhysRevD.80.083507}{{\em Phys.Rev.}
  {\bfseries D80} (2009) 083507},
\href{http://arxiv.org/abs/0907.0705}{{\ttfamily arXiv:0907.0705
  [astro-ph.CO]}}.

\bibitem{Dai:2013kfa}
{\sc L.~Dai}, {\sc D.~Jeong}, {\sc M.~Kamionkowski}, and {\sc J.~Chluba},
  ``{The Pesky Power Asymmetry},''
\href{http://arxiv.org/abs/1303.6949}{{\ttfamily arXiv:1303.6949
  [astro-ph.CO]}}.

\bibitem{Hirata:2013sb}
{\sc C.~M. Hirata}, ``The cmb power asymmetry,'' 2013.

\bibitem{Green:2009ds}
{\sc D.~Green}, {\sc B.~Horn}, {\sc L.~Senatore}, and {\sc E.~Silverstein},
  ``{Trapped Inflation},''
  \href{http://dx.doi.org/10.1103/PhysRevD.80.063533}{{\em Phys.Rev.}
  {\bfseries D80} (2009) 063533},
\href{http://arxiv.org/abs/0902.1006}{{\ttfamily arXiv:0902.1006 [hep-th]}}.

\bibitem{LopezNacir:2011kk}
{\sc D.~Lopez~Nacir}, {\sc R.~A. Porto}, {\sc L.~Senatore}, and {\sc
  M.~Zaldarriaga}, ``{Dissipative effects in the Effective Field Theory of
  Inflation},'' \href{http://dx.doi.org/10.1007/JHEP01(2012)075}{{\em JHEP}
  {\bfseries 1201} (2012) 075},
\href{http://arxiv.org/abs/1109.4192}{{\ttfamily arXiv:1109.4192 [hep-th]}}.

\bibitem{DAmico:2012sz}
{\sc G.~D'Amico}, {\sc R.~Gobbetti}, {\sc M.~Kleban}, and {\sc M.~Schillo},
  ``{Inflation from Flux Cascades},''
\href{http://arxiv.org/abs/1211.3416}{{\ttfamily arXiv:1211.3416 [hep-th]}}.

\bibitem{DAmico:2012ji}
{\sc G.~D'Amico}, {\sc R.~Gobbetti}, {\sc M.~Kleban}, and {\sc M.~Schillo},
  ``{Unwinding Inflation},''
  \href{http://dx.doi.org/10.1088/1475-7516/2013/03/004}{{\em JCAP} {\bfseries
  1303} (2013) 004},
\href{http://arxiv.org/abs/1211.4589}{{\ttfamily arXiv:1211.4589 [hep-th]}}.

\bibitem{Hirata:2009ar}
{\sc C.~M. Hirata}, ``{Constraints on cosmic hemispherical power anomalies from
  quasars},'' \href{http://dx.doi.org/10.1088/1475-7516/2009/09/011}{{\em JCAP}
  {\bfseries 0909} (2009) 011},
\href{http://arxiv.org/abs/0907.0703}{{\ttfamily arXiv:0907.0703
  [astro-ph.CO]}}.

\bibitem{Lyth:2013vha}
{\sc D.~H. Lyth}, ``{The CMB asymmetry from inflation},''
\href{http://arxiv.org/abs/1304.1270}{{\ttfamily arXiv:1304.1270
  [astro-ph.CO]}}.

\bibitem{Liddle:2013rta}
{\sc A.~R. Liddle} and {\sc M.~Corts}, ``{Cosmic microwave background
  anomalies in an open universe},''
\href{http://arxiv.org/abs/1306.5698}{{\ttfamily arXiv:1306.5698
  [astro-ph.CO]}}.

\bibitem{Alishahiha:2004eh}
{\sc M.~Alishahiha}, {\sc E.~Silverstein}, and {\sc D.~Tong}, ``{DBI in the
  sky},'' \href{http://dx.doi.org/10.1103/PhysRevD.70.123505}{{\em Phys.Rev.}
  {\bfseries D70} (2004) 123505},
\href{http://arxiv.org/abs/hep-th/0404084}{{\ttfamily arXiv:hep-th/0404084
  [hep-th]}}.

\bibitem{Gordon:2005ai}
{\sc C.~Gordon}, {\sc W.~Hu}, {\sc D.~Huterer}, and {\sc T.~M. Crawford},
  ``{Spontaneous isotropy breaking: a mechanism for cmb multipole
  alignments},'' \href{http://dx.doi.org/10.1103/PhysRevD.72.103002}{{\em
  Phys.Rev.} {\bfseries D72} (2005) 103002},
\href{http://arxiv.org/abs/astro-ph/0509301}{{\ttfamily arXiv:astro-ph/0509301
  [astro-ph]}}.

\bibitem{Kofman:2004yc}
{\sc L.~Kofman}, {\sc A.~D. Linde}, {\sc X.~Liu}, {\sc A.~Maloney}, {\sc
  L.~McAllister}, et~al., ``{Beauty is attractive: Moduli trapping at enhanced
  symmetry points},''
  \href{http://dx.doi.org/10.1088/1126-6708/2004/05/030}{{\em JHEP} {\bfseries
  0405} (2004) 030},
\href{http://arxiv.org/abs/hep-th/0403001}{{\ttfamily arXiv:hep-th/0403001
  [hep-th]}}.

\bibitem{Kofman:1997yn}
{\sc L.~Kofman}, {\sc A.~D. Linde}, and {\sc A.~A. Starobinsky}, ``{Towards the
  theory of reheating after inflation},''
  \href{http://dx.doi.org/10.1103/PhysRevD.56.3258}{{\em Phys.Rev.} {\bfseries
  D56} (1997) 3258--3295},
\href{http://arxiv.org/abs/hep-ph/9704452}{{\ttfamily arXiv:hep-ph/9704452
  [hep-ph]}}.

\bibitem{Dvorkin:2007jp}
{\sc C.~Dvorkin}, {\sc H.~V. Peiris}, and {\sc W.~Hu}, ``{Testable polarization
  predictions for models of CMB isotropy anomalies},''
  \href{http://dx.doi.org/10.1103/PhysRevD.77.063008}{{\em Phys.Rev.}
  {\bfseries D77} (2008) 063008},
\href{http://arxiv.org/abs/0711.2321}{{\ttfamily arXiv:0711.2321 [astro-ph]}}.

\bibitem{GZ}
{\sc L.~P. {Grishchuk}} and {\sc I.~B. {Zeldovich}}, ``{Long-wavelength
  perturbations of a Friedmann universe, and anisotropy of the microwave
  background radiation},'' {\em Sov. Astron.} {\bfseries 22} (Apr., 1978)
  125--129.

\bibitem{Kleban:2011pg}
{\sc M.~Kleban}, ``{Cosmic Bubble Collisions},''
  \href{http://dx.doi.org/10.1088/0264-9381/28/20/204008}{{\em
  Class.Quant.Grav.} {\bfseries 28} (2011) 204008},
\href{http://arxiv.org/abs/1107.2593}{{\ttfamily arXiv:1107.2593
  [astro-ph.CO]}}.

\bibitem{Kleban:2011yc}
{\sc M.~Kleban}, {\sc T.~S. Levi}, and {\sc K.~Sigurdson}, ``{Observing the
  Multiverse with Cosmic Wakes},''
\href{http://arxiv.org/abs/1109.3473}{{\ttfamily arXiv:1109.3473
  [astro-ph.CO]}}.

\bibitem{Gobbetti:2012yq}
{\sc R.~Gobbetti} and {\sc M.~Kleban}, ``{Analyzing Cosmic Bubble
  Collisions},'' \href{http://dx.doi.org/10.1088/1475-7516/2012/05/025}{{\em
  JCAP} {\bfseries 1205} (2012) 025},
\href{http://arxiv.org/abs/1201.6380}{{\ttfamily arXiv:1201.6380 [hep-th]}}.

\end{thebibliography}\endgroup

\end{document}